\documentclass[12pt, letterpaper]{article}
\usepackage[utf8]{inputenc}
\usepackage[english]{babel}
\usepackage{times}
\usepackage{geometry}
\usepackage{fancyhdr}
\usepackage{enumerate}
\usepackage{enumitem,amssymb}
\usepackage{pifont}
\usepackage{blindtext}
\usepackage{graphicx}
\usepackage{xcolor}
\usepackage{soul}
\usepackage[compact]{titlesec}
\usepackage[colorlinks]{hyperref}
\usepackage{doi}
\usepackage{wrapfig}
\usepackage{subcaption}

\usepackage{amsthm,amssymb,mathptmx}

 \hypersetup{
     colorlinks=true,
     linkcolor=magenta,
     filecolor=blue,
     citecolor=cyan,      
     urlcolor=blue,
     }

\newcommand\farcsec{\mbox{$.\mkern-4mu^{\prime\prime\mkern-2mu}$}}%
\newcommand\arcsec{\mbox{$^{\prime\prime}$}}%

\begin{document}
\thispagestyle{empty}
\newlength{\oldparindent}
\setlength{\oldparindent}{\parindent}
\setlength{\parindent}{0cm}
\noindent

\begin{center}
{\Large\textbf{Development of Integral Field Spectrographs \\ to Revolutionize Spectroscopic Observations \\ of Solar Flares and other Energetic Solar Eruptions \\}} \bigskip
{\large\textbf{\textit{White Paper for}} \vspace{0.5ex}\\}
{\large\textbf{2022 Solar and Space Physics (Heliophysics) Decadal Survey}}
\end{center}

\begin{center}
    \begin{tabular}{rl}
    \textbf{Category:} & Basic Research \\
    \textbf{Topic:} & Solar Physics
    \end{tabular}
\end{center}

\subsubsection*{Principal Author}
\begin{tabular}{@{}ll}
Name: & Dr. Haosheng Lin ({\small\texttt{haosheng@hawaii.edu}}) \\	
Affiliation: & Institute for Astronomy, University of Hawaii \\  
\end{tabular}

\subsubsection*{Co-authors}
Tetsu Anan (National Solar Observatory),
Gianna Cauzzi (National Solar Observatory),
Lyndsay Fletcher (University of Glasgow, University of Oslo), 
Pei Huang (Ball Aerospace),
Adam Kowalski (University of Colorado),
Maxim Kramar (Institute for Astronomy, University of Hawaii),
Jiong Qiu (Montana State University),
Jenna Samra (Center for Astrophysics $|$ Harvard \& Smithsonian),
Constance Spittler (Ball Aerospace),
Takashi Sukegawa (Canon, Inc., Japan), 
Gregory Wirth (Ball Aerospace)

\subsubsection*{Co-signers}
Hugh Hudson (University of Glasgow),
Ryoko Ishikawa (National Astronomical Observatory of Japan), 
Yukio Katsukawa (National Astronomical Observatory of Japan),
Yoshinori Suematsu (National Astronomical Observatory of Japan)

\vspace*{\fill}
\subsection*{Abstract}
The Sun's proximity offers us a unique opportunity to study in detail the physical processes on a star's surface; however, the highly dynamic nature of the stellar surface -- in particular, energetic eruptions such as flares and coronal mass ejections -- presents tremendous observational challenges. Spectroscopy probes the physical state of the solar atmosphere, but conventional scanning spectrographs and spectrometers are unable to capture the full evolutionary history of these dynamic events with a sufficiently wide field of view and high spatial, spectral, and temporal resolution.    \textbf{\textit{Resolving the physics of the dynamic sun requires gathering simultaneous spectra across a contiguous area over the full duration of these events, a goal now tantalizingly close to achievable with continued investment in developing powerful new Integral Field Spectrographs to serve as the foundation of both future ground- and space-based missions.}}  This technology promises to revolutionize our ability to study solar flares and CMEs, addressing NASA's strategic objective to ``understand the Sun, solar system, and universe.'' Since such events generate electromagnetic radiation and high-energy particles that disrupt terrestrial electric infrastructure, this investment not only advances humanity's scientific endeavors but also enhances our space weather forecasting capability to protect against threats to our technology-reliant civilization.

\newpage
\setlength{\parindent}{\oldparindent}
\section{Key Science Goals}

Flares, coronal mass ejections (CMEs), and other energetic solar phenomena like filament eruptions and penumbral jets are long-standing enigmas of solar physics. 
Resolving what physical mechanisms trigger flares, heat the solar atmosphere, and accelerate the resulting energetic particles requires quantitative measurements of the magnetic and thermodynamic properties of the solar atmosphere at flare sites and surrounding regions before, during, and after flares. Measurements of the magnetic field, temperature, and density of the solar atmosphere during such dynamic events can only be achieved through rapid spectroscopic and polarimetric observations of spectral lines sensitive to the local magnetic and thermodynamic conditions. 

\paragraph{Science Case 1: Flare Spectroscopy.} Flares are the result of dynamic magnetic field reconfiguration above the stellar surface which accelerates particles and heats the surrounding atmosphere. Broadband spectroscopy of the hydrogen Balmer line series is 
one of the most powerful diagnostic tools for  identifying the mechanisms heating the stellar atmosphere during flares \cite{Kowalski2017Broadening}. The Balmer continuum strength constrains the depth of the heating by non-thermal particles, the broadening and series merging of the Balmer lines in the blue indicates the ambient electron density, and the spectra of the H$\alpha$, $\beta$ and $\gamma$ lines constrain the NLTE (non-local thermal equilibrium), opacity broadening effects higher in the chromosphere \cite{2015ApJ...813..125K, Kowalski2022}. 
Blueshifts in the hydrogen Balmer series observed in other stars are typically interpreted as signatures of filament eruptions and CMEs \cite{Vida2019}, but the connection to corresponding spectral phenomena on the Sun remains uncertain \cite{Canfield1990}.
Finally, H$\alpha$ linear impact polarization could diagnose non-thermal particle energy and angular distributions \cite{2003ASPC..307..480H, 2013A&A...556A..95H}. 
No other observations provide such a comprehensive set of diagnostics. 

\begin{figure}[htb]
\centering
\includegraphics[scale=0.32]{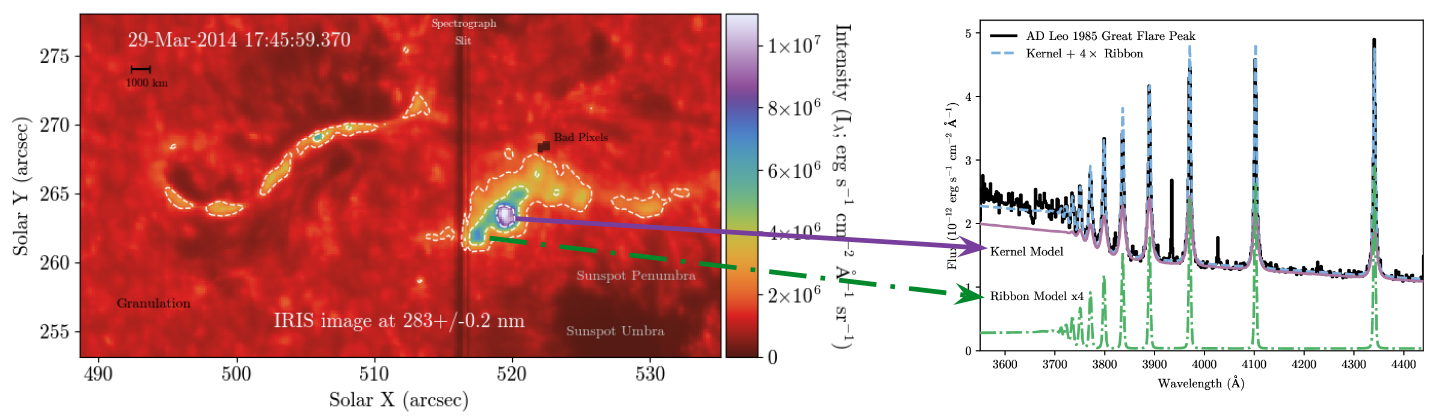}
\vspace{-0.25cm}
\caption{
    \textbf{(Left)} Spatially resolved solar flare images from the IRIS spacecraft and \textbf{(Right)} stellar ``superflare'' spectra from \cite{HP91}.  The images show both circular kernels and more diffuse ribbons.  Integral field spectroscopy can probe whether these two structures are consistent with models derived from observations of stellar flares, in which a small Balmer jump and strong optical continuum originate from the smaller kernels, and the emission line flux (Balmer series, Ca~II) predominantly originates from the diffuse ribbons. 
} 
\label{fig:adleo}
\end{figure}

Figure~\ref{fig:adleo} (left panel) illustrates the appearance of solar flare ribbons. Such spatially resolved observations of solar flares reveal that different parts of the flare evolve in distinct ways. However, few existing spectral observations have sufficient spatial resolution to capture the state of the solar atmosphere in different parts (kernels or ribbons) of the flare. On the other hand, many broadband spectroscopic observations of powerful stellar flares have revealed large variations in the characteristics of the hydrogen Balmer lines during flare events. The right panel of Fig.~\ref{fig:adleo} shows the spectrum of a ``super flare'' event on the nearby red dwarf star AD Leonis. This spatially unresolved spectrum is modeled as a simple combination of a flare kernel source and flare ribbons. 

The 
thus-far unpredictable nature of solar flares has made spectroscopic flare observations extremely challenging. Flare-producing active regions can usually be identified in advance \cite{2016ApJ...829...89B}, but the precise location of the highly intermittent flare kernels and ribbons is unknown \textit{a priori} (cf. Fig.~\ref{fig:adleo}), causing instruments with a limited field of view (FOV) frequently to miss the critical impulsive phase at the onset of the flare. 
Thus, optimal flare observations require both high resolution and extended coverage in the spatial, spectral, and temporal domains. Because of these challenges and the limitations of existing instruments at blue wavelengths, \textbf{\textit{not a single spectroscopic observation covering the entire evolution of a solar flare exists to date.}} Even with the unprecedented successes of the Interface Region Imaging Spectrograph (IRIS) in observing flare ribbon dynamics, the longslit rastering and sit-and-stare modes have often missed the brightest kernels 
(see  Fig.~\ref{fig:adleo}, left) \cite{Li_2017}. Far more severe ambiguities of slit placement relative to the bright kernels existed in the pioneering work of the 1980s \cite{Ichimoto1984, Neidig1983}. Progress requires new instrumentation able to capture spectra over a large 2D region simultaneously with sufficient spatial, spectral, \textit{and} temporal resolution across a wide wavelength range containing most of the hydrogen Balmer lines.  

\paragraph{Science Case 2: Coronal Magnetometry.} Measuring coronal magnetic fields \cite{2000ApJ...541L..83L,LinKuhnCoulter2004} also requires spectra covering both space and time. White light and extreme ultraviolet (EUV) spectral imaging of the solar corona reveals a host of fascinating coronal structures and dynamics shaped and controlled by the magnetic field, including
energetic eruptions that eject plasma at both high temperature (coronal mass ejections) and low temperature (filament eruptions). The triggering and accelerating mechanisms, as well as the physical connection between CMEs and flares, are unclear \cite{2003ApJ...586..562G, 2005JGRA..11012S05Y}, but given the expected low plasma $\beta$ of the inner corona, magnetic fields are accepted as the dominant force, with reorganization of the coronal magnetic fields providing the energy.
Observations of the 3D coronal magnetic and thermodynamic structures before, during, and after CME events will illuminate the underlying physical processes. 
Knowledge of the global 3D coronal magnetic field will also improve heliosphere models and space weather forecasting.

The low density, high temperature, and weak magnetic fields in the corona conspire to make measuring the coronal magnetic fields exceptionally challenging.  Only a handful of diagnostic tools exists for directly measuring the coronal magnetic fields, including spectropolarimetry of the coronal emission lines (CELs) in the visible and IR \cite{1998ApJ...500.1009J, 1999ApJ...522..524C, 2000ApJ...542..528L}, coronal seismology \cite{2007Sci...317.1192T, 2020Sci...369..694Y}, gyrosynchrotron radiation at radio wavelengths \cite{1997ApJ...488..488B, 1999AstL...25..250B, 2006ApJ...641L..69B}, and most recently the magnetic induced transition (MIT) effect of EUV spectral lines \cite{2021ApJ...913..135L, 2021ApJ...915L..24B}, each with their strengths and limitations. For example, the polarized spectra of CELs contain information about the transverse direction and longitudinal field strength of the magnetic fields, but are limited to off-limb observations for visible and infrared CELs.  Conversely, while the MIT effect directly measures the magnetic field strength, it does not provide information on the direction of the magnetic field. Most notable and frustrating is that due to the low opacity of the coronal plasma, coronal measurements are sensitive to contributions from all structures along the line of sight (LOS) and cannot be interpreted as indicators of the local physical condition. Only tomographic observations that involve simultaneous observations of an object from multiple nonredundant view points and employ tomographic inversion techniques to reconstruct the 3D structure of the observed object can disentangle the LOS integration effect to reveal the 3D structure of the corona \cite{2006A&A...456..665K, 2013ApJ...775...25K, 2016ApJ...819L..36K, 2016FrASS...3...25K}. 

The development of new instrumentation and interpretation tools over the past two decades has succeeded in measuring polarization in CELs, providing regular observations of the direction of the transverse component in coronal magnetic fields \cite{2019ApJ...883...74K}. However, direct measurement of the \textit{strength} of the coronal magnetic field measurements has not been demonstrated with tunable filter based spectrographs; \textbf{\textit{integral field spectropolarimetry remains the only proven method for directly 
mapping the strength of coronal magnetic fields}}. While the newly commissioned Daniel K. Inouye Solar Telescope (DKIST) is poised to provide more frequent and higher precision coronal magnetic field measurements \cite{2020SoPh..295..172R}, the small FOV resulting from its relatively large aperture limits its ability to make efficient global-scale observations. 

\begin{wrapfigure}{l}{0.5\textwidth} 
\centering{\includegraphics[width=0.5\textwidth]{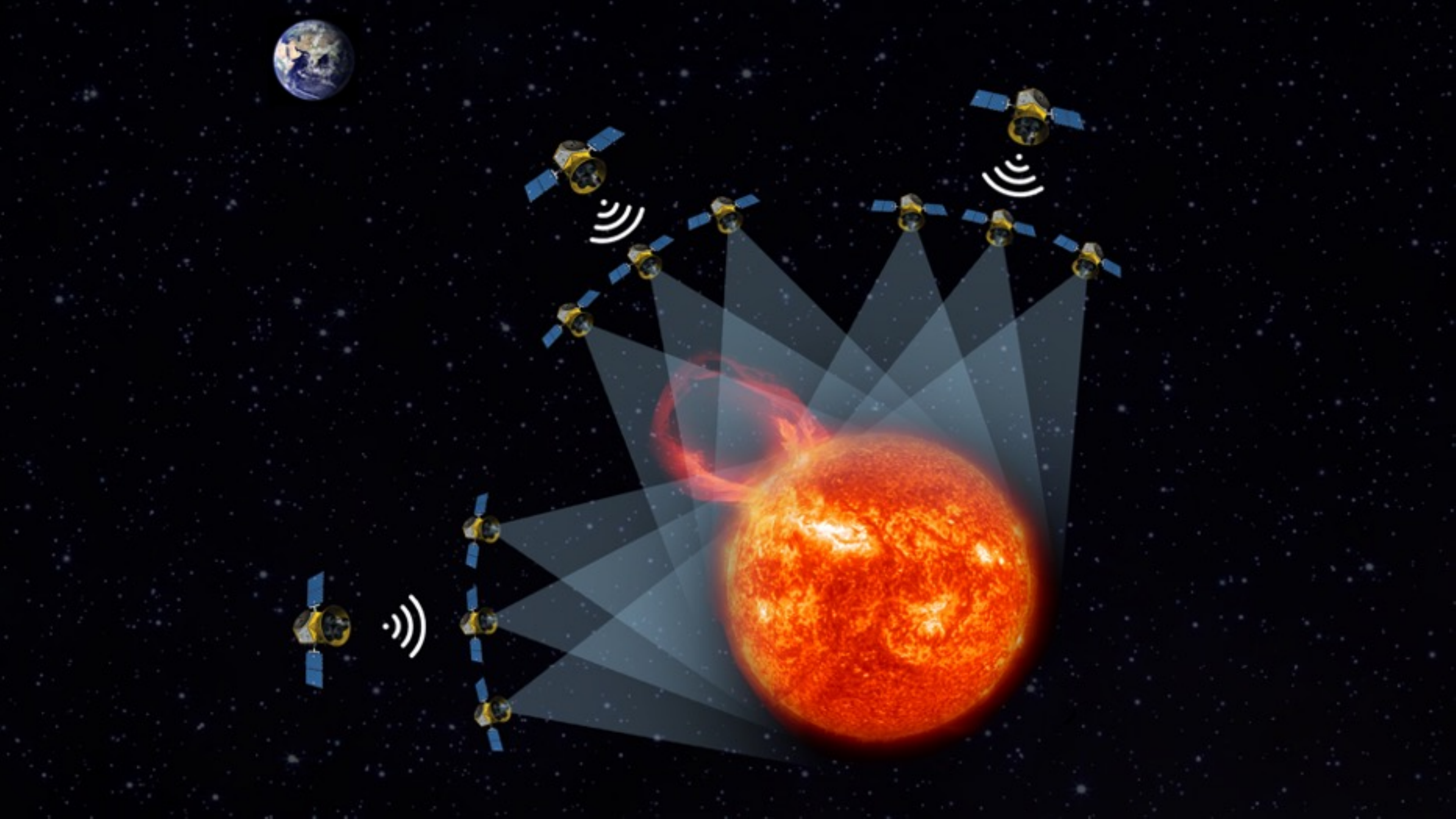}}
\caption{
    Mission concept for the Solar TOmogRaphic Magnetometry (STORM) Mission, using multiple spacecraft to simultaneously acquire multi-viewpoint measurements of the solar coronal magnetic fields with IFS coronal spectropolarimeters.  Vector tomographic inversion can reconstruct the 3D magnetic, temperature, and density structure of the corona during CMEs. 
}
\label{Fig:SolarSTORM}
\vspace*{-1em}
\end{wrapfigure}

Understanding CMEs requires large FOV, high-temporal-resolution observations spanning the entire evolutionary history of the events. Wide-field, spatially resolved measurements of coronal magnetic fields are possible with the aid of an Integral Field Spectrograph (IFS), as clearly demonstrated by the early work with fiber-optic-based IFSs  \cite{LinKuhnCoulter2004}. New medium-resolution (5\arcsec--10\arcsec) coronal spectropolarimeters equipped with advanced IFSs from ground- and space-based observatories with global coverage (0.5 -- 1 $R_{\odot}$) will enable direct observational validation of coronal models \cite{2008ApJ...680.1496L,2014ApJ...787L...3R} to advance coronal magnetic field research in conjunction with  high-resolution DKIST and space EUV observations.
To push further, constellations of spacecraft equipped with EUV imagers, IFSs, coronal spectropolarimeters, and photospheric vector magnetometers  in circumsolar orbits observing from multiple view points (Fig.~\ref{Fig:SolarSTORM}) will eventually enable tomographic reconstruction of the dynamic 3D magnetic and thermodynamic structures of the solar corona. For example, magnetic free energy (MFE) is believed to be the source of energy powering CMEs. Probing the 3D coronal magnetic field structure via tomography will directly measure the MFE of the eruption corona before, during, and after CMEs \cite{2021ApJ...907...23C}.


\section{Technical Overview and Technology Drivers}
This section highlights an ongoing effort to demonstrate Flare Sentinel, an Integral Field Spectrograph targeting spectroscopic observation of the hydrogen Balmer series spectral lines during solar flares. This demonstration also serves to mature the technology and build confidence for future development of IFS for coronal spectropolarimetry and other applications.

\begin{wrapfigure}{r}{0.5\textwidth}
\centering{\includegraphics[width=0.45\textwidth]{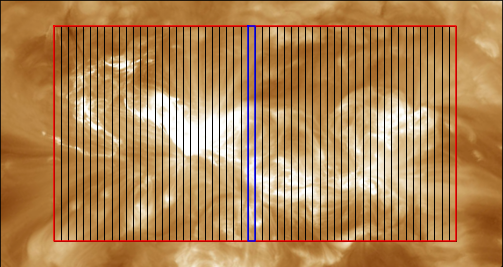}}
\caption{Comparison of observed regions for traditional vs.\ Integral Field Spectrographs.  Traditional slit spectrographs only capture a spectrum within a thin region (shown in blue), requiring them to ``scan'' the slit to capture the spectrum across a region of interest (outlined in red).  In contrast, the image-slicer-based IFS in the DL-NIRSP instrument on DKIST \cite{2021AAS...23810602W} uses 56 ``slits'' (outlined in black) to  monitor the full region of interest simultaneously. \textit{(Image credit: NASA/GSFC)}
\label{Fig:slicer-FOV}}
\end{wrapfigure}

\paragraph{Technological options for Integral Field Spectroscopy.}
The principal hurdle to observing dynamic solar events is the inability of traditional spectrographs to gather simultaneous spectra covering the full region of interest.  A conventional slit spectrograph can only observe a tiny region at any moment, so capturing spectra over a larger region requires field scanning perpendicular to the slit in a series of exposures (Fig.~\ref{Fig:slicer-FOV}).  Alternatively, slitless imaging spectrographs (e.g., Fourier-transform spectrographs and tunable Fabry-Perot filters) must employ wavelength scanning to capture 2D spectra over a fixed region of interest \cite{CRISP2008, iFTS2014}. 
Due to their inability to observe all wavelengths in a 2D region at once, neither design is suitable for studying rapidly-evolving phenomena such as solar flares and CMEs.

In contrast, an IFS can overcome these limitations to acquire 
3D hyperspectral data cubes with high temporal/spatial/spectral resolution.  
A standard IFS consists of two parts: an Integral Field Unit (IFU) that reformats a 2D spatial field into an 1D linear field, and a grating spectrograph to disperse the light. 
IFUs come in several different types (lenslet arrays, optical fiber arrays, and free-space image slicers) which each feature particular advantages and limitations when employed for heliophysics observations.  \textit{Lenslet arrays}, consisting of a 2D array of refractive lenses placed $<$1~mm apart on a single substrate \cite{10.1117/12.672061}, can typically observe only a short wavelength range and make inefficient use of space on the detector \cite{1995A&AS..113..347B}.  Contemporary \textit{ fiber arrays} can achieve a high ``fill factor'' in the focal plane and provide excellent throughput \cite{10.1117/12.2230740,10.1117/12.2311289}, but offer limited spectral coverage, potentially high cross-coupling between adjacent channels, and susceptibility to temperature/pressure swings and radiation effects in spaceborne applications. Certain fiber materials also suffer from reversible darkening under intense illumination, making them poorly suited for solar observations \cite{Neveux:93}.

\begin{wrapfigure}{l}{0.6\textwidth}
\centering{\fbox{\includegraphics[width=0.525\textwidth]{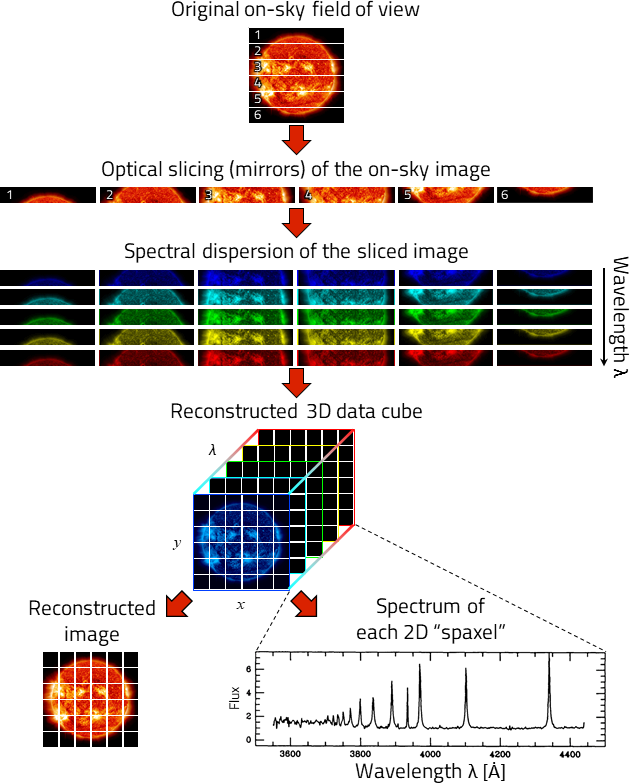}}}
\caption{
    Operation of an image slicer style IFS.  The region of interest is imaged onto the ``slicer,'' which lies in the focal plane and consists of many thin mirrors.  The slicer re-images the scene into multiple slices aligned on the dispersing element, generating a separate spectrum for each slice.  The resulting spectra form a 3D ``data cube'' which captures a spectrum at each point within the 2D region of interest simultaneously. \textit{(Figure per \cite{JWST-IFS}; Spectrum per \cite{HP91}; Image credit: NASA/SDO)}
\label{Fig:slicer}}
\vspace{-3ex}
\end{wrapfigure}

These and other limitations of lenslet and fiber IFSs prompted development of \textit{free-space image slicers} that can be employed in ground- and space-based solar telescopes. An image slicer consists of a series of thin mirrors that reformats a 2D scene into a series of slits, allowing a downstream dispersive element to create a spectrum of each point in the scene (Fig.~\ref{Fig:slicer}).  Image-slicer-based IFSs are highly stable, highly efficient, and can cover a wide wavelength range.  The principal shortcoming of existing diamond turned and polished glass image slicers has been the difficulty in fabricating slicer mirrors with widths under 100~$\mu$m. Because the spectral resolution of a spectrograph varies with the width of the slicer mirrors (equivalent to the slit of the spectrograph), most existing image slicer IFSs exhibit low spectral resolution. High-resolution instruments have traditionally required  large spectrographs and have thus  been unsuitable for space applications \cite{GRIS2012,GRIS2016}.

\begin{figure}[h]
\begin{center}
\includegraphics[height=11cm]{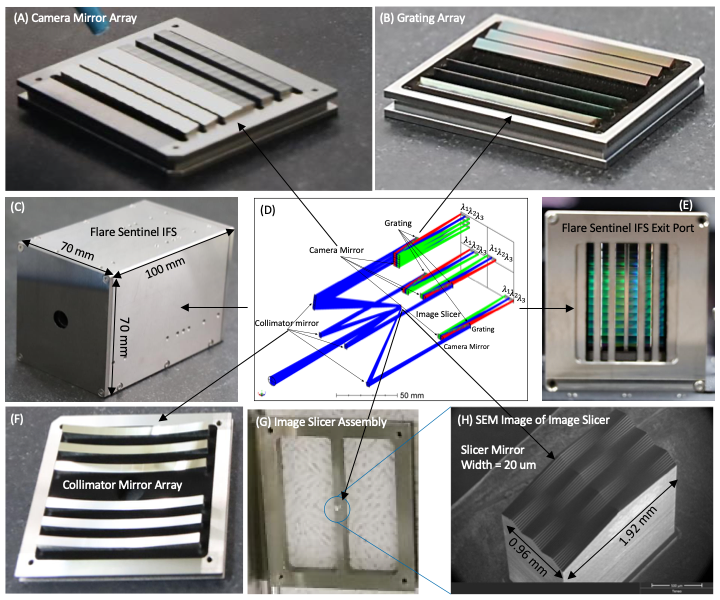}
\caption{
    Optical design (Panel D) and pictures of a new compact IFS prototype assembly and its major optical components demonstrating the manufacturability of the new IFS design. The IFS is designed for a $54~{\rm mm}\times40~{\rm mm}$ sensor and covers the 3500--4500~\AA\ spectral window, encompassing the hydrogen Balmer continuum and the Balmer series up to the H$\gamma$ line with $R=2,000$ spectral resolution. The beams for 3500, 4000, and 4500~\AA\ are shown in green, blue, and red, respectively. This IFS provides a $48\times96\times2000$ ($n_x, n_y, n_{\lambda}$) format hyperspectral data cube in one exposure. The entire spectrograph resides in a compact $70~{\rm mm}\times70~{\rm mm}\times100~{\rm mm}$ volume as shown in Panel C. All components were fabricated on Invar substrates, making the system mechanically robust for space missions. 
\label{Fig:MSA_Parts}
}
\vspace{-22pt}
\end{center}
\end{figure}

\paragraph{State of the art.}
Recent advancements in free-form \textit{diamond cutting} optics machining technology by Canon, Inc., in Japan \cite{2014SPIE.9151E..1SS} have introduced image-slicer IFUs with sub-100~$\mu$m slicer mirrors. Spectrographs built with this technology offer the high spatial and spectral resolution, high efficiency, and low scattered light levels required for solar observations, as demonstrated by the prototype IFU for the Diffraction-Limited Near IR Spectropolarimeter on DKIST \cite{2022SPIE.12188E..78SS}. Figure~\ref{Fig:MSA_Parts} shows the optical design of an innovative compact IFS, plus pictures of the mechanical assembly and major components containing the active optical elements of a prototype that Canon recently fabricated successfully. This IFS consists of a machined image slicer block followed by an array of miniature spectrographs, as shown in the Zemax ray tracing (panel D). Panel C shows the compact spectrograph. Panel E is a view of the IFS exit port showing the dispersed spectra. 

The centerpiece of this new IFS design is the small monolithic image slicer block with 96 slicer mirrors each $20~{\mu}{\rm m}\times0.92~{\rm mm}$  in size, arranged in a $48\times2$ configuration located at the center of the image (see circle in Panel G). A scanning electron microscope image of the image slicer block appears in panel H. \textbf{\textit{The ability to fabricate 20~$\mu$m slicer mirrors represents a major technology breakthrough in the field of imaging spectroscopy.}} Doing so allowed us to incorporate an array of miniature spectrographs into the IFU, eliminating the need for a large spectrograph in the conventional IFS design and yielding an exceptionally compact IFS without compromising optical performance. Panels A and F show the camera mirror array and collimator mirror array, respectively. Each component contains 96 small powered mirrors fabricated on a monolithic substrate. The grating array (panel B) consists of 96 surface-relieved gratings, also fabricated on a monolithic substrate. The successful fabrication of this challenging prototype demonstrates the feasibility of this new IFS design. 

\begin{figure}[h]
\begin{center}
\fbox{\includegraphics[height=5cm]{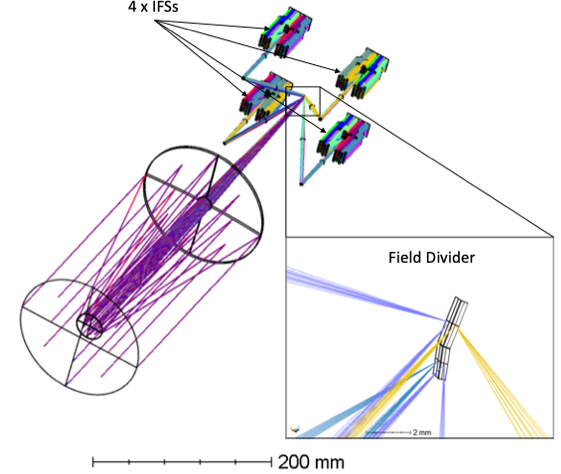}}
\fbox{\includegraphics[height=5cm]{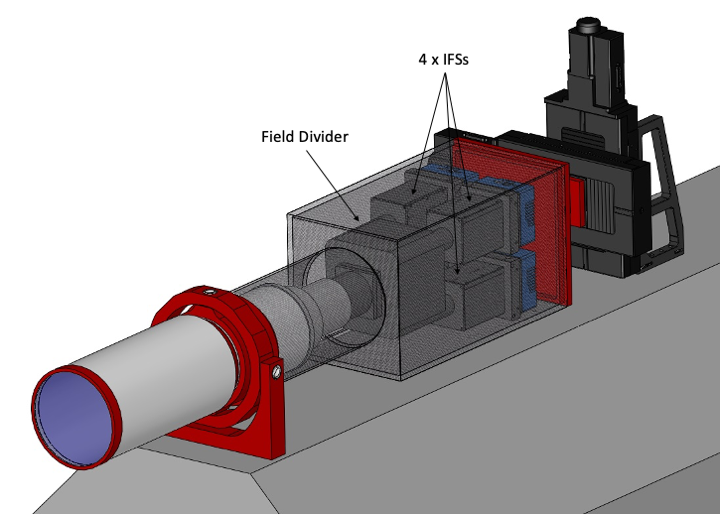}}
\caption{
    \textbf{(Left)} Optical design of a telescope equipped with 4 identical IFSs. A field divider (lower right corner insert) separates the field into four beams for simultaneous observation by the IFSs. 
    \textbf{(Right)} Conceptual design of Flare Sentinel telescope and spectrometer equipped with 4 IFSs.
    \label{Fig:MSA_OpticalDesign}
}
\vspace{-22pt}
\end{center}
\end{figure}

\paragraph{Flare Sentinel --- a technology demonstration.}
The IFS design of Fig.~\ref{Fig:MSA_Parts} is highly flexible and can be optimized to meet the requirements of wide-ranging applications. Its compactness also makes it easily scalable. For example, Fig.~\ref{Fig:MSA_OpticalDesign} shows the conceptual design of Flare Sentinel, an imaging spectrograph with $96\times192\times2000$ ($n_x, n_y, n_{\lambda}$) hyperspectral format equipped with four IFSs for flare spectroscopy of hydrogen Balmer series spectral lines. Recent observations of solar white-light flares indicate that the size of the hottest kernels can be less than $0\farcsec5$ \cite{2019ApJ...878..135K,2020ApJ...895....6G}. Each IFS of Flare Sentinel has a $29\arcsec\times58\arcsec$ FOV with $0\farcsec6$/pixel spatial sampling, giving Flare Sentinel a combined $58\arcsec\times116\arcsec$ FOV that can \textbf{\textit{fully cover a typical active solar region in a single pointing}}. This large field size will yield complete spatial coverage of large flares (GOES class M, or X). Broadband UV/optical imaging and hard X-ray observations have clearly shown that the energy release processes during flares operate at a temporal scale of 1~s or shorter \cite{2012A&A...547A..72Q}. The best large-field flare spectra to date were obtained with cadence of on the order of 10--30~s \cite{2015ApJ...813..125K, 2018ApJ...860...10K}. 
Using modern high-speed cameras, Flare Sentinel can achieve a temporal resolution of 10~Hz or higher, limited only by the frame rate of the camera. Flare Sentinel will finally open the door to time-resolved 2D spectroscopy of dynamic solar events with complete coverage of their evolution. 

\section{Connection to Heliophysics Decadal Survey Objectives}


The charter for the present Decadal Survey for Solar and Space Physics indicates that the report will address two science areas relevant to the present work: 
\begin{enumerate}[nolistsep]

    \item \textit{The structure of the Sun and the properties of its outer layers in their static and active states.} IFS technology will open new pathways to the study of active regions on the sun and thus pertains directly to this science area.    
    
    \item \textit{Science related to the interstellar medium, astrospheres (including their stars), exoplanets, and planetary habitability.} With their ability to collect spectra and images for a contiguous region centered on a star hosting a planetary system while simultaneously monitoring wavefront quality \cite{Pope_2014}, IFSs are ideally suited to exoplanet studies such as those envisioned with a large UV-optical-IR space telescope as proposed in the Astro2020 Decadal Survey \cite{NAP26141}.

\end{enumerate}
In summary, progress in developing IFS technology will impact not only solar astronomy but also exoplanet science, two areas of keen interest in heliophysics.

\section{Conclusion}

More than a century of observations have led to a comprehensive understand of the \textit{static} sun; however, our current capabilities are still insufficient to decipher the physics of the \textit{dynamic} sun. As Fletcher \textit{et al.} \cite{2010arXiv1011.4650F} lamented in a White Paper submitted to the 2010 Heliophysics Decadal Survey: \textit{``It is an embarrassment that there is still no imaging spectroscopy of solar flares from space.''} 
While IRIS has partially remedied this shortcoming in the UV \cite{De_Pontieu_2014}, advancements in optics and large format focal plane array fabrication technologies have finally made the many advantages of wide-FOV integral field spectroscopy from UV to far-IR feasible from space.
%
Putting instruments such as Flare Sentinel in space will enable round-the-clock monitoring and observations to trace the evolution of thermodynamic properties of solar flares.

And yet, flares are just one aspect of the dynamic sun. Achieving a comprehensive understanding of the energetic solar eruptions that can occur at every spatial scale requires employing all the observational tools at our disposal simultaneously. IFS technologies have now reached a critical point, with their potential for \textit{\textbf{improving observing efficiency by almost two orders of magnitude}} over conventional slit spectrographs finally putting revolutionary scientific advancements within our grasp. The roadmap below outlines steps needed to advance the technology that will lead to an eventual STORM mission.  Timely investment in this area is needed to further advance IFS fabrication technologies, encourage more innovative instrument designs, and advance their Technology Readiness Level to lay a solid foundation for their deployment in space missions in the near future.  

\begin{figure}[htb]
\begin{center}
\includegraphics[width=0.95\textwidth]{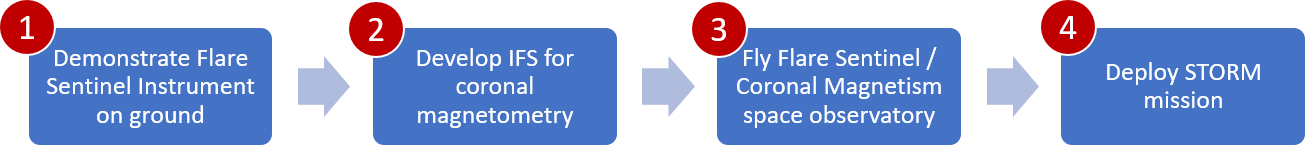}
\end{center}
\end{figure}

\newpage
\bibliographystyle{IEEEtranDOI} 
\bibliography{biblioflareCME}

\end{document}